\documentclass[conference]{IEEEtran}
\pdfoutput=1 
\IEEEoverridecommandlockouts
\usepackage{cite}
\usepackage{amsmath,amssymb,amsfonts}
\usepackage{algorithmic}
\usepackage[pdftex]{graphicx}
\usepackage{textcomp}
\usepackage{multirow}
\usepackage[pdftex]{algorithm}
\usepackage{flushend}
\usepackage[pdftex]{color}
\usepackage{soul}
\usepackage[pdftex]{xcolor}
\usepackage[pdftex]{hyperref}

\usepackage{fancyhdr}
\fancypagestyle{firstpage}{% Page style for first page
  \fancyhf{}% Clear header/footer
  \fancyhead[C]{\small Extended version of the paper ``DFTS2: Deep Feature Transmission Simulation for Collaborative Intelligence,'' to be presented at the \textit{IEEE Visual Communications and Image Processing (VCIP)}, Munich, Germany, December 5-8, 2021.}
}

\begin{document}
\title{\vspace*{0.5cm}DFTS2: Simulating Deep Feature Transmission Over Packet Loss Channels\\
\thanks{This work was supported in part by the NSERC grant RGPIN-2021-02485.}
}

\author{\IEEEauthorblockN{Ashiv Dhondea, Robert A. Cohen, and Ivan V. Baji\'{c}}
\IEEEauthorblockA{\textit{School of Engineering Science, Simon Fraser University, Burnaby, BC, V5A 1S6, Canada}} %\\
}

\maketitle

\begin{abstract}
In edge-cloud collaborative intelligence (CI), an unreliable transmission channel exists in the information path of the AI model performing the inference. It is important to be able to simulate the performance of the CI system across an imperfect channel in order to understand system behavior and develop appropriate error control strategies. In this paper we present a simulation framework called DFTS2, which enables researchers to define the components of the CI system in TensorFlow~2, select a packet-based channel model with various parameters, and simulate system behavior under various channel conditions and error/loss control strategies. Using DFTS2, we also present the most comprehensive study to date of the packet loss concealment methods for collaborative image classification models.  
\end{abstract}


\thispagestyle{firstpage}

\section{Introduction}
\label{sec:introduction}
Collaborative intelligence (CI)~\cite{neurosurgeon,CI_Overview_ICASSP2021} is a ``hybrid" AI deployment strategy. To complete an AI task, CI typically involves splitting the Deep Neural Network (DNN) computations between an edge device and a cloud service. The resource-constrained edge device runs the initial few layers of the DNN on the input signal (image, video, ...) and produces a deep feature tensor. The latter is transmitted over a communication channel to the cloud, where the remaining layers of the DNN complete the inference task. CI has shown potential for improved inference latency and energy use as compared to purely cloud-based or edge-based AI deployment approaches in certain scenarios~\cite{neurosurgeon,jointdnn}.

Potential applications of CI, which include IoT systems, traffic monitoring, autonomous vehicles or drones and point-of-care systems, may involve imperfect communication channels. This motivates the need to investigate the performance of the CI system across such channels in order to understand system behavior and develop error/loss resilience strategies. To this end, DFTS -- Deep Feature Transmission Simulator~\cite{unnibhavi2018dfts} -- was developed to study the effects of missing feature data on inference accuracy in CI. Here we present a more sophisticated simulation framework called DFTS2, with several new capabilities: additional channel models and missing feature recovery methods from the recent literature.

This paper is organized as follows. Section~\ref{sec:related_work} discusses related work on packet loss concealment in CI. Section~\ref{sec:description} describes the operation of the proposed simulation framework. Experiments are discussed in Section~\ref{sec:experiments} followed by conclusions in Section~\ref{sec:conclusions}. An implementation of DFTS2 is available online, along with packet traces and simulation files for results presented in this work.\footnote{\url{https://github.com/AshivDhondea/DFTS2}}

\section{Related work} \label{sec:related_work}
Two approaches to error resilience in CI have been studied in the literature: 1) joint source-channel coding of deep features~\cite{choi_neural_2019,BottleNet++} and 2) packet-based feature transmission with simple error concealment, such as bilinear  and nearest-neighbor interpolation in~\cite{unnibhavi2018dfts}, tensor completion~\cite{Bragile2020,CALTeC_ICIP_2021}, and inpainting~\cite{Bajic2021objdet}.

Generic tensor completion methods such as Simple Low Rank Tensor Completion (SiLRTC)~\cite{liu2012tensor} and High Accuracy Low Rank Tensor Completion (HaLRTC)~\cite{liu2012tensor} were adapted to recover corrupted deep feature data packets in~\cite{Bragile2020,CALTeC_ICIP_2021,Bajic2021objdet}. Furthermore,~\cite{Bragile2020} proposed a method called ALTeC - Adaptive Linear Tensor Completion - specifically designed to recover missing intermediate features by pre-training on deep feature tensors. With a single-feature-row-per-packet scheme over a random loss channel, ALTeC proved to be as least as accurate as SiLRTC and HaLRTC while bringing a significant reduction in inference latency~\cite{Bragile2020}. In~\cite{CALTeC_ICIP_2021}, a Content Adaptive Linear Tensor Completion (CALTeC) method was proposed to exploit existing redundancies in deep feature tensors. With a multiple-feature-rows-per-packet scheme over a Gilbert-Elliott channel, CALTeC and HaLRTC produced similar Top-1 classofication accuracy in~\cite{CALTeC_ICIP_2021}. ALTeC slightly outperformed CALTeC and HaLRTC in one scenario and offered slightly worse performance than the other two in another scenario when the split point of the DNN was closer to its input and the number of feature rows per packet was increased. SiLRTC-completed tensors led to the poorest Top-1 classification accuracy in both scenarios in~\cite{CALTeC_ICIP_2021}. In~\cite{Bajic2021objdet}, PDE-based image inpainting methods,  ``Navier-Stokes"~\cite{navierstokes} and ``Telea"~\cite{telea2004image}, performed lost packet concealment in collaborative object detection. Both inpainting methods led to higher mean average precision than SiLRTC with 250 iterations, while being more than 100$\times$ faster. 

This paper presents an extended version of DFTS~\cite{unnibhavi2018dfts}, called DFTS2, with more advanced capabilities for simulating packet loss and loss recovery. Specifically, the Gilbert-Elliot channel model~\cite{5755057} and packet traces are introduced. Moreover, the recent feature recovery mechanisms mentioned above - SiLRTC, HaLRTC, ALTeC, CALTeC, and feature inpainting - have been implemented. Using DFTS2, we present a comprehensive comparison of the mentioned feature recovery methods on collaborative image classification.

\section{DFTS2 simulator description}
\label{sec:description}
\begin{figure}
    \centering
    \includegraphics[width=0.48\textwidth]{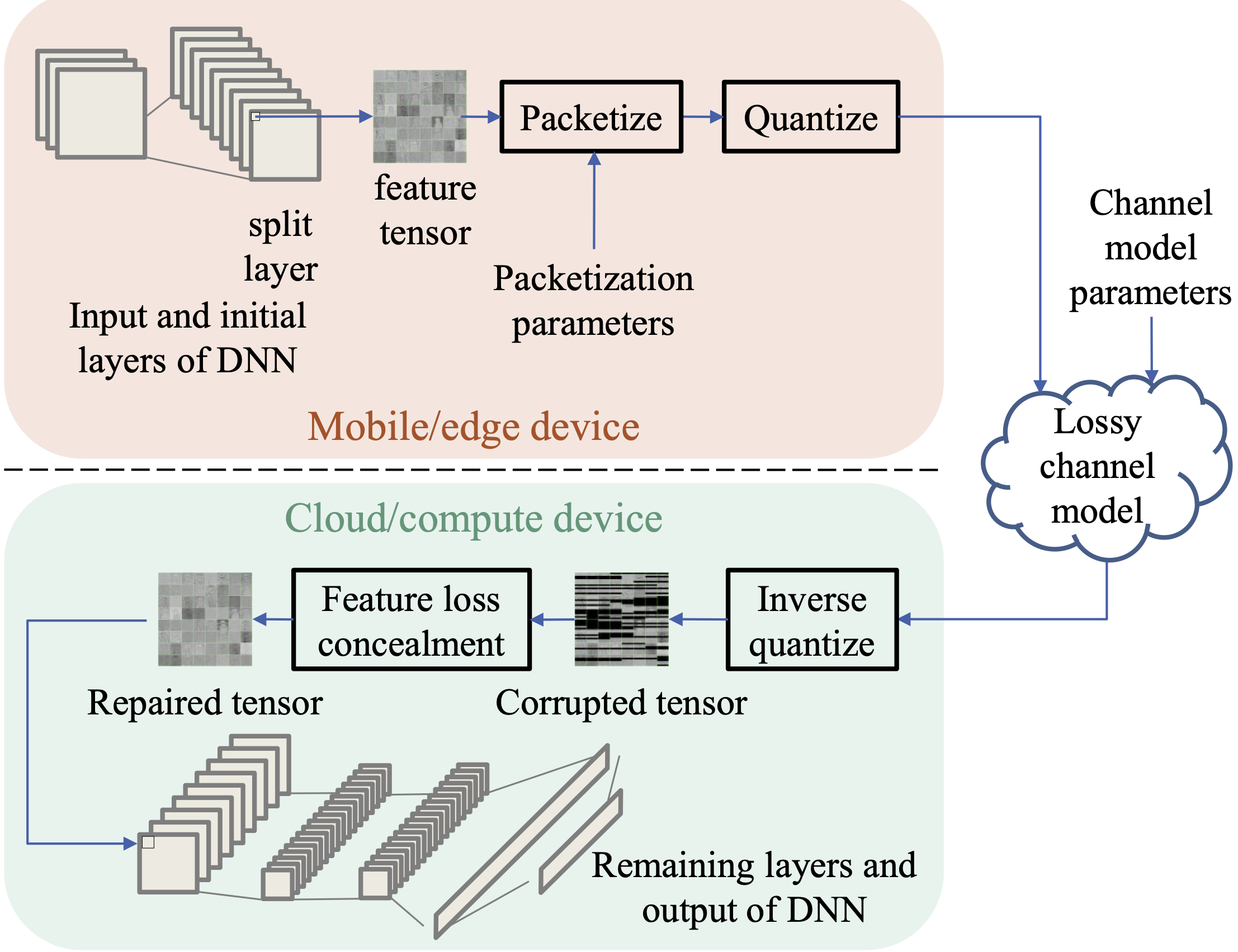}
    \caption{DFTS2 system overview (Adapted from~\cite[Fig. 1]{cohen2021lightweight})}
    \label{fig:description:systemoverview}
\end{figure}

DFTS2 splits TensorFlow 2 DNN models into a \textit{mobile} (or \textit{edge}) sub-model and a \textit{cloud} (or \textit{remote}) sub-model at a user-defined split point. After processing an input image, the mobile sub-model produces an intermediate feature tensor, which is transmitted over an imperfect communication channel to the remote service. Prior to transmission over a packet network, the feature tensor is packetized and quantized. Fig.~\ref{fig:description:systemoverview} gives a system-level overview of a DFTS2 experiment. A lossy communication channel between the mobile device and the remote service leads to data packets being damaged or lost due to errors in data transmission and network congestion, for example. The different components involved in a DFTS2 experiment are summarized in Table~\ref{table:description:components} and discussed in detail in the following subsections.

%The following subsections discuss packetization, communication channel model and packet loss concealment in DFTS2.
%the main components involved in tensor transmission from the edge to the cloud.

\subsection{Intermediate feature tensor packetization} \label{subsec:description:pkt}
\begin{figure}
\centering
\includegraphics[width=7.4cm]{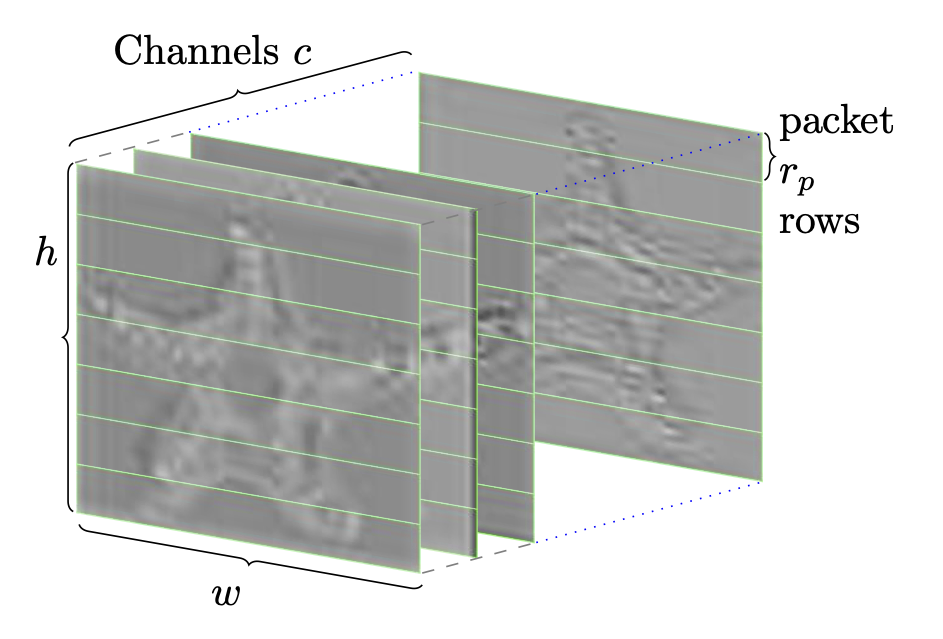}
    \caption{A tensor $\mathcal{X} \in \mathbb{R}^{h \times w \times c}$ from layer \texttt{add\_1} of ResNet-18. Several consecutive rows in each channel form a feature data packet.}
    \label{fig:tensorviz}
\end{figure}

A feature tensor with height $h$, width $w$, and the number of channels $c$ will be denoted $\mathcal{X} \in \mathbb{R}^{h \times w \times c}$. An illustration of a feature tensor extracted from layer \texttt{add\_1} of ResNet\nobreakdash-18~\cite{ResNet} is shown in Fig.~\ref{fig:tensorviz}.
Layer \texttt{add\_1} corresponds to the output of the shortcut connection and element-wise addition applied to the final output of the \texttt{conv2\_x} layers, as denoted in~\cite[Table~1]{ResNet}. Currently, there is no ``standard" approach to packetizing deep feature tensor data. DFTS2 employs a packetization scheme inspired by a common practice in video streaming, where video frames are encoded into packets of consecutive rows of frame elements~\cite{video2002}.
%\hl{Is the term "frame elements" used in that book? If not, then perhaps saying "packets of consecutive rows" is sufficient.}
Similarly, each channel in a DFTS2 tensor is partitioned into groups of $r_p$ rows, as indicated by the dashed
%\hl{remove the word "dashed?"}
green lines in Fig.~\ref{fig:tensorviz}. Zero padding is done on the last packet of a channel if the number of channel rows ($h$) is not an integer multiple of the number of packet rows ($r_p$). The user-defined $r_p$ was set to 8 rows in our experiments. Packetized tensors were min-max quantized~\cite{choi2018deep} to 8 bits/element. While it is possible to further compress the data by additional prediction, quantization, entropy coding~\cite{Hyomin_MMSP_2018,Saeed_ICIP_2019,BaF_ICASSP_2021,lightweight_ICME_2020}, and/or bit allocation~\cite{bit_alloc_TIP_2021}, this was not done in our experiments because the focus is on error resilience, not on compression.

%The $i$-th packet in the $j$-th tensor channel is denoted $\mathbf{X}_i^{(j)} \in \mathbb{R}^{r_p \times w}$, for $j \in \left\{0, 1,  \dotsc, c-1\right\}$. The vectorized version of the packet is denoted $\mathbf{x}_i^{(j)} = \text{Vec}\left(\mathbf{X}_i^{(j)}\right) \in \mathbb{R}^{(r_p\cdot w) \times 1}$. In our simulations, the tensors were min-max quantized~\cite{choi2018deep} to 8 bits/element. Further compression of the data in each packet is possible, but since our focus here is not on compression, no further quantization or entropy coding is used.  

\subsection{Channel models} \label{subsec:description:channel}
Four options for the communication channel model are offered in DFTS2 experiments: perfect channel (no loss), independent and identically distributed (iid) random loss, Gilbert-Elliott (GE) channel~\cite{5755057}, and user-provided packet traces. The GE model, also known as the burst loss model, has been shown to adequately capture observed packet loss patterns over the Internet~\cite{5755057,stuhlmuller2000analysis}. This motivates our choice of GE channel models for our experiments. This packet loss model is a two-state Markov model with a good $(G)$ and a bad $(B)$ state. It can be parameterized by the burst loss probability $P_B$ and the average burst length $L_B$. The conversion from $P_B$ and $L_B$ to the four state transition probabilities is as follows:
\begin{equation}
\begin{alignedat}{2}
    p_{B\to G} & = \frac{1}{L_B},  \quad && p_{G\to B} = \frac{P_B}{L_B(1-P_B)},\\
    p_{B\to B} & =1-p_{B\to G}, \quad
    && p_{G\to G} = 1-p_{G\to B}.
\end{alignedat}
\end{equation} 
Since DFTS2 can also load packet traces from imported files to create corrupted deep feature tensor packets, it allows the user to provide packet traces from an external source, such as from actual measurements over a network or from a communication network simulation package.

When a feature tensor is transmitted, some of its packets will be lost, depending on the channel realization. The remote client identifies missing packets from packet sequence numbers provided by a transport-layer protocol and reconstitutes the available part of the tensor. The available data is inverse quantized, and the missing data (identified by a Boolean loss map) is recovered using an error concealment method. 

\subsection{Packet loss concealment} \label{subsec:description:concealment}
Missing value recovery in corrupted deep feature tensors means that more accurate tensor data can be exploited by the back-end sub-model, resulting in significant gains in classification accuracy~\cite{CALTeC_ICIP_2021}. 

SiLRTC~\cite{liu2012tensor} and HaLRTC~\cite{liu2012tensor} assume that the original, un-corrupted tensors are low-rank. They iteratively reduce the rank of the damaged tensor via singular value shrinkage to try to recover the original tensor values. Due to their iterative operation, these general tensor completion methods introduce significant processing delays, which is a problem in applications sensitive to inference latency. SiLRTC and HaLRTC make no assumption about the spatial distribution of missing values in a corrupted tensor, or how the tensor was created in the first place. %as long as the indices of the correct tensor values are known. Indices for the correct values are derived from the Boolean loss maps. 
In~\cite{Bragile2020}, the Frobenius norm of the difference between completed tensors over successive iterations of tensor completion with SiLRTC and HaLRTC was found to have reached a plateau within 50 iterations. Therefore, we ran SiLRTC and HaLRTC for 50 iterations in our experiments.

Besides SiLRTC and HaLRTC, DFTS2 also includes ALTeC~\cite{Bragile2020}, CALTeC~\cite{CALTeC_ICIP_2021}, and Navier-Stokes inpainting~\cite{navierstokes}. All these methods exploit spatial structure of feature tensor channels to recover missing data. ALTeC assumes a linear relationship between a packet of feature data and its spatial neighbors, as well as collocated data in other feature channels. The coefficients of this linear relationship are obtained by pre-training. Meanwhile, for each missing feature packet, CALTeC finds the most correlated feature channel where collocated data is available, and then derives an affine transformation to fill in the missing data. Navier-Stokes inpainting is based on a ``surface flow'' of available spatially neighboring data into the missing region.
%In~\cite{CALTeC_ICIP_2021}, ALTeC was adapted to operate on the multiple-feature-rows-per-packet scheme adopted in DFTS2. ALTeC was pre-trained on 5000 images from the ILSVRC 2012 validation set~\cite{imagenet2015} on inverse-quantized deep feature tensors for each specific DNN and split layer. An image inpainting-based method, ``Navier-Stokes" \footnote{\url{https://docs.opencv.org/4.5.2/df/d3d/tutorial_py_inpainting.html}}~\cite{navierstokes} can also be used for packet loss concealment. Inpainting masks are obtained from channel loss maps. Finally, CALTeC, which requires neither pre-training nor parameter tuning, was also used for tensor completion.

\subsection{Simulation mode}
The simulator can run in a ``single-shot'' mode or a Monte Carlo simulation mode. In the single-shot mode, a single image is input to the edge sub-model, the transmission of the corresponding feature packets is simulated, and the received features are processed by the cloud sub-model. Monte Carlo simulation operates like a single-shot mode, but in batches of images, and over multiple channel realizations, so that average performance can be measured.  

\begin{table}[t]   \caption{Main simulator components}
  %\vspace{4pt}
  \label{table:description:components}
  \centering
 \begin{tabular}{ l | l }
 %\hline
   \textbf{Component} & \textbf{Options} \\
      \hline
      \hline
      \textbf{Model} & Any TensorFlow 2 DNN model \\
      \hline  
      \textbf{Split layer} & Any layer \\
      \hline
      \multirow{2}{*}{\textbf{Quantization}} & No quantization \\  & Uniform $n$-bit \\
      \hline
      \textbf{Packetization} & $r_p$ feature rows per packet \\
      \hline
      \multirow{3}{*}{\textbf{Channel model}} & Independent \& identically distributed (iid)\\
       & Gilbert-Elliott\\
       & Packet traces    \\
      \hline
      \multirow{6}{*}{\textbf{Feature loss concealment}}  & None \\
       & SiLRTC~\cite{liu2012tensor} \\
       & HaLRTC~\cite{liu2012tensor} \\
       & ALTeC~\cite{Bragile2020} \\
       & CALTeC~\cite{CALTeC_ICIP_2021} \\
       & Navier-Stokes~\cite{navierstokes}\\
      \hline
      \multirow{2}{*}{\textbf{Simulation mode}} & Single-shot \\
       & Monte Carlo\\
      \hline
    \end{tabular}
\end{table}

\section{Experiments} 
\label{sec:experiments}
Here we use DFTS2 to conduct what we believe are the most comprehensive experiments on packet loss resilient collaborative image classification. Experiments  were run on the same 882 images used in~\cite{unnibhavi2018dfts,CALTeC_ICIP_2021}, which are a subset of 10 classes of the ImageNet~\cite{imagenet2015} test set. Pre-trained\footnote{\url{https://github.com/qubvel/classification_models}} DNNs, namely ResNet\nobreakdash-18~\cite{ResNet}, ResNet\nobreakdash-34~\cite{ResNet}, DenseNet\nobreakdash-121~\cite{densenet2017}, and EfficientNet\nobreakdash-B0~\cite{efficientnetpaper}, were each split at the output of an early layer with a single connection to downstream layers. 
This simulation scenario corresponds to a mobile/edge device having very limited computational capabilities,
%which can execute a minimal amount of computation,
hence the split at an early layer. Of course, DFTS2 also supports split layers deeper in the DNN. %With a single downstream connection, the amount of transmitted data to be minimized for reduced inference latency.
Layer \texttt{add\_1} for ResNet\nobreakdash-18 and layer \texttt{add\_3} for ResNet\nobreakdash-34 correspond to the output of the shortcut connection and element-wise addition applied to the final output of the \texttt{conv2\_x} layers~\cite[Table~1]{ResNet}. This means that a single deep feature tensor
%$\mathcal{X} \in 56 \times 56 \times 64$
$\mathcal{X} \in \mathbb{R}^{56 \times 56 \times 64}$
is to be transmitted %from the edge device 
to the cloud.
For DenseNet\nobreakdash-121, \texttt{pool2\_conv} is the output of convolution layer in the first transition layer block, as denoted in~\cite[Table~1]{densenet2017}. In this case, $\mathcal{X} \in \mathbb{R}^{56 \times 56 \times 128}$. In the case of EfficientNet\nobreakdash-B0 \texttt{block2b\_add} tensors, $\mathcal{X} \in  \mathbb{R}^{56 \times 56 \times 24}$. With $r_p = 8$ rows of features/packet, we end up with 7 packets per tensor channel in the case of ResNet\nobreakdash-18 \texttt{add\_1}, ResNet\nobreakdash-34 \texttt{add\_3}, DenseNet\nobreakdash-121 \texttt{pool2\_conv}, and EfficientNet\nobreakdash-B0 \texttt{block2b\_add}. When ResNet\nobreakdash-18 is split at the output of the \texttt{add\_3} layer, the transmitted feature tensor has dimensions $28 \times 28 \times 64$. With $r_p = 4$ rows of features/packet, we end up with 7 packets per tensor channel.

% The tensor $\mathcal{X}$ has dimensions $56 \times 56 \times 128$ for the DenseNet\nobreakdash-121 case. 
%For the DenseNet\nobreakdash-121 case, $\mathcal{X} \in \mathbb{R}^{56 \times 56 \times 128}$.
Tensors are min-max quantized to $n=8$ bits before transmission over a communication channel.  

DFTS2 experiments were run using Python 3.6 and TensorFlow~2 in this study. All packet concealment methods employed are executed on CPU only, though it may be possible to develop GPU-compatible implementations.

\begin{figure}
\centering
\includegraphics[width=7.4cm]{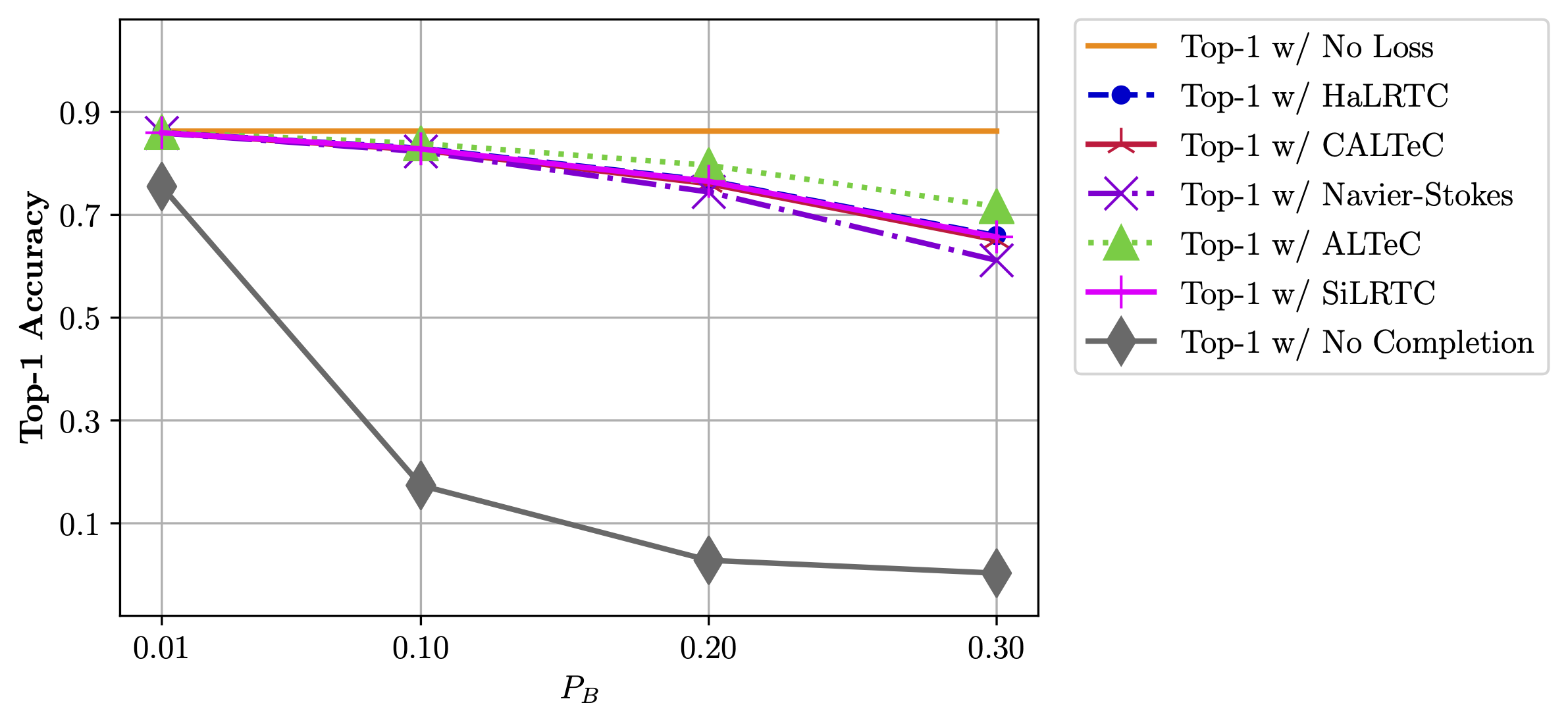}
    \caption{Top-1 cloud sub-model classification accuracy averaged over Monte Carlo simulations under a range of GE burst lengths on EfficientNet-B0 \texttt{block2b\_add} tensors.}
    \label{fig:efficientnetb0}
\end{figure}

\begin{figure*}[tb]
 \center
		\begin{minipage}[b]{0.24\linewidth}
		\centering
		\includegraphics[width=4cm]{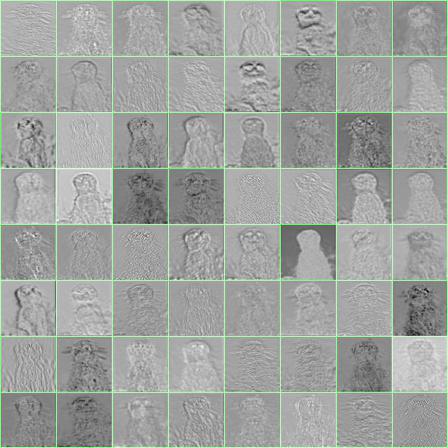}
		(a) Original tensor
	\end{minipage}
\begin{minipage}[b]{0.24\linewidth}
	\centering
	\includegraphics[width=4cm]{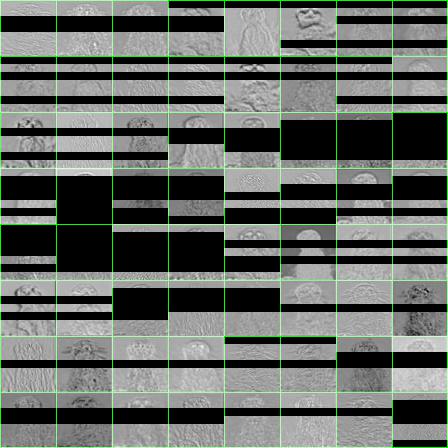}
	(b) Corrupted tensor
\end{minipage}
\begin{minipage}[b]{0.24\linewidth}
	\centering
	\includegraphics[width=4cm]{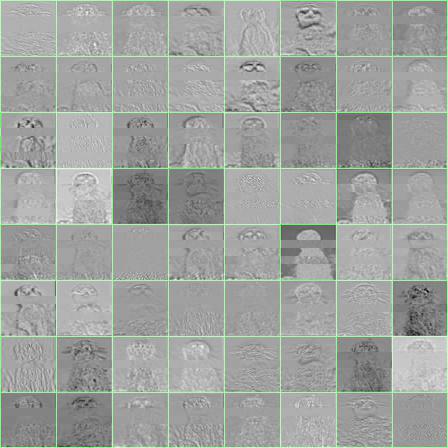}
	(c) CALTeC-repaired tensor
	\end{minipage}
\begin{minipage}[b]{0.24\linewidth}
	\centering
	\includegraphics[width=4cm]{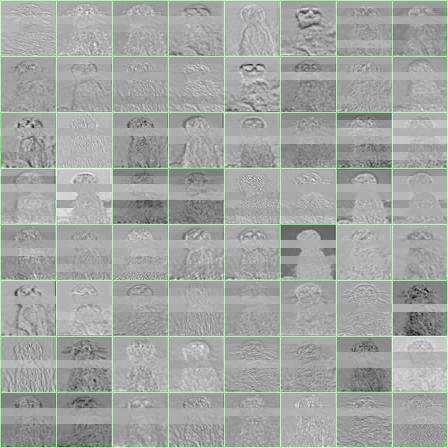}
	(d) ALTeC-repaired tensor
\end{minipage}
\hfill
\vspace{6pt}
	\begin{minipage}[b]{0.24\linewidth}
		\centering
		\includegraphics[width=4cm]{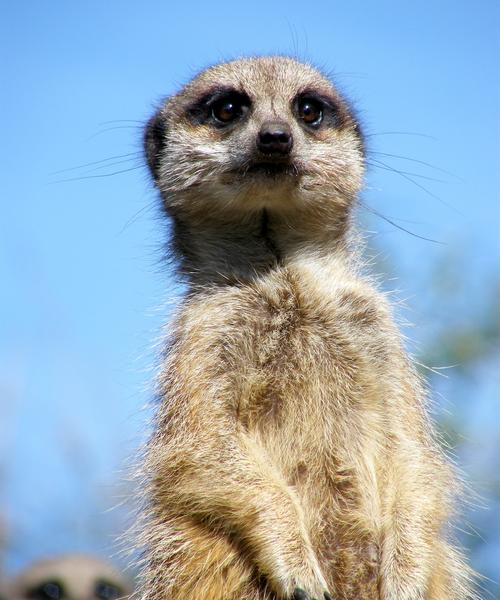}
		(h) Meerkat test image
	\end{minipage}
\begin{minipage}[b]{0.24\linewidth}
	\centering
	\includegraphics[width=4cm]{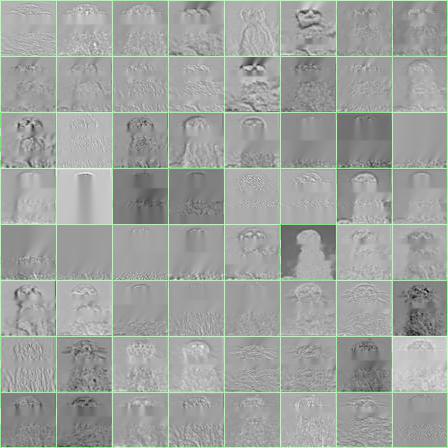}
	(g) Navier-Stokes inpainting
\end{minipage}
\begin{minipage}[b]{0.24\linewidth}
	\centering
	\includegraphics[width=4cm]{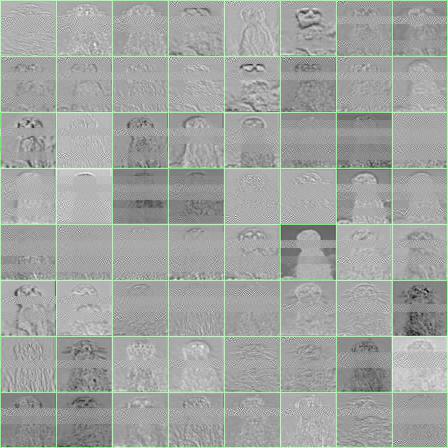}
	(f) SiLRTC-50
	\end{minipage}
\begin{minipage}[b]{0.24\linewidth}
	\centering
	\includegraphics[width=4cm]{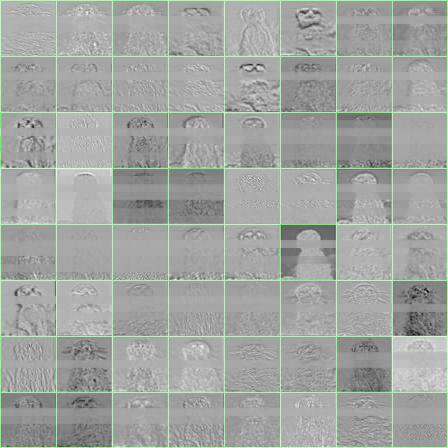}
	(e) HaLRTC-50
	\end{minipage}
\caption{Tiled ResNet-34 add\_3 tensor produced from a (h) Meerkat image: (a) original; (b) corrupted; repaired with (c) CALTeC, (d) ALTeC, (e) HaLRTC with 50 iterations, (f) SiLRTC with 50 iterations and (g) Navier-Stokes inpainting. Each tile delineated by green lines represents a channel in a feature tensor. Images were mapped to grayscale for enhanced visualization.}
\label{fig:exp:tiled}
\end{figure*}

Packets were transmitted over GE channels with burst loss probabilities $P_B \in \{0.01, 0.10, 0.20, 0.30\}$ and average burst lengths of $L_B \in \{1, 2, \dotsc, 7\}$. %These can be converted to GE transition probabilities as explained in Section~\ref{subsec:description:channel}. 
The range of burst loss probabilities chosen for Monte Carlo experiments is in accordance with a conventional practice in video streaming research~\cite{video2002}. 
%With 8 rows of tensor elements per packet, each $56 \times 56$ tensor channel in our experiments is thus packetized into 7 packets ($7\cdot8=56$).
All tensor channels in our experiments are packetized into 7 packets: $56 \times 56$ tensor channels with 8 rows of tensor elements per packet ($7\cdot8=56$) and $28 \times 28$ tensor channels with 4 rows of features per packet ($7\cdot4=28$). 
The range of $L_B$ was chosen up to 7, to cover the case where the entire tensor channel is likely to be lost. In our Monte Carlo experiments, for each pair of $(P_B,L_B)$ values, 20 channels realizations were simulated for each batch of the 882 image test set. Boolean loss maps for each test batch over 20 channel realizations of each GE channel $(P_B,L_B)$ were saved. These loss maps were then used to re-create packet loss patterns so that the same corrupted tensors were presented to each loss concealment method.

In the results that follow, `No Loss' or `NL' is the performance of the system without packet loss or quantization, i.e., the performance of the baseline DNN model without splitting into edge and cloud sub-models. `No Completion'or `NC' means no tensor completion, that is, tensor elements from missing packets are simply set to zero. For each loss probability $P_B \in \{0.01,0.10,0.20,0.30\}$, the cloud sub-model Top-1 accuracy performance was accumulated over all average burst lengths and averaged. Tables \ref{table:resnet18add1}, \ref{table:resnet18add3}, \ref{table:resnet34}, \ref{table:densenet121pool2conv}, and \ref{table:efficientnetb0} summarize Monte Carlo results for all simulation scenarios considered. Fig.~\ref{fig:efficientnetb0} shows the Top-1 cloud sub-model classification accuracy over Monte Carlo experiments for EfficientNet\nobreakdash-B0 \texttt{block2b\_add} tensors. Each marker in Fig.~\ref{fig:efficientnetb0} thus represents 140 Monte Carlo runs through the 882 image test set. For all DNNs, SiLRTC and HaLRTC were run for 50 iterations, which were found to be sufficient for convergence in~\cite{Bragile2020}. Feature recovery done with all packet loss concealment methods leads to significant improvements in Top-1 classification accuracy over the `No Completion\ case in Fig.~\ref{fig:efficientnetb0}.

\begin{table}[t]
\centering
\caption{ResNet\nobreakdash-18 \texttt{add\_1} tensors: classification accuracy averaged over 140 Monte Carlo runs per burst loss probability $P_B$. `NL' refers to the `No Loss' case; `NC' refers to the `No Completion' case; other entries refer to the cloud prediction accuracy on tensors repaired with the indicated method. The highest classification accuracy values in each scenario are indicated in bold. }
\label{table:resnet18add1}
\begin{tabular}{|l|llll|}
\hline
$\boldsymbol{P_B}$  & \textbf{0.3}      & \textbf{0.2}      & \textbf{0.1}      & \textbf{0.01}     \\
\hline
\hline
\textbf{Top-1 NL}            & 85.37\% & 85.37\% & 85.37\% & 85.37\% \\
\textbf{Top-5 NL}            & 94.44\% & 94.44\% & 94.44\% & 94.44\% \\
\hline
\textbf{Top-1 NC}            & 34.07\% & 55.42\% & 74.85\% & 84.70\% \\
\textbf{Top-5 NC}            & 49.70\% & 72.21\% & 88.26\% & 94.17\% \\
\hline
\textbf{Top-1 ALTeC}         & 58.57\% & 75.75\% & 83.05\% & 85.31\% \\
\textbf{Top-5 ALTeC}         & 74.41\% & 88.17\% & 93.15\% & \textbf{94.43\%} \\
\hline
\textbf{Top-1 CALTeC}        & \textbf{69.71\%} & 78.81\% & 83.30\% & \textbf{85.33\%} \\
\textbf{Top-5 CALTeC}        & \textbf{84.54\%} & \textbf{91.04\%} & 93.64\% & 94.43\% \\
\hline
\textbf{Top-1 HaLRTC}        & 68.77\% & \textbf{78.82\%} & \textbf{83.51\%} & 85.33\% \\
\textbf{Top-5 HaLRTC}        & 83.54\% & 90.94\% & \textbf{93.70\%} & 94.43\% \\
\hline
\textbf{Top-1 SiLRTC}        & 57.43\% & 67.21\% & 73.55\% & 82.66\% \\
\textbf{Top-5 SiLRTC}        & 70.50\% & 78.74\% & 84.00\% & 92.47\% \\
\hline
\textbf{Top-1 Navier-Stokes} & 58.81\% & 74.12\% & 81.86\% & 85.20\% \\
\textbf{Top-5 Navier-Stokes} & 75.85\% & 88.09\% & 92.97\% & 94.41\% \\
\hline
\end{tabular}
\end{table}

\begin{table}[t]
\centering
\caption{ResNet\nobreakdash-18 \texttt{add\_3} tensors: classification accuracy averaged over 140 Monte Carlo runs per burst loss probability $P_B$}
\label{table:resnet18add3}
\begin{tabular}{|l|llll|}
\hline
$\boldsymbol{P_B}$                & \textbf{0.3}      & \textbf{0.2}      & \textbf{0.1}      & \textbf{0.01}               \\
\hline
\hline
\textbf{Top-1 NL}            & 85.37\% & 85.37\% & 85.37\% & 85.37\%           \\
\textbf{Top-5 NL}            & 94.44\% & 94.44\% & 94.44\% & 94.44\%           \\
\hline
\textbf{Top-1 NC}            & 65.67\% & 75.46\% & 81.62\% & 85.11\%           \\
\textbf{Top-5 NC}            & 82.82\% & 89.32\% & 92.68\% & 94.36\%           \\
\hline
\textbf{Top-1 ALTeC}         & \textbf{81.62\%} & \textbf{84.01\%} & \textbf{85.16\%} & \textbf{85.31\%}           \\
\textbf{Top-5 ALTeC}         & \textbf{92.66\%} & \textbf{93.87\%} & \textbf{94.29\%} & \textbf{94.49\%}           \\
\hline
\textbf{Top-1 CALTeC}        & 76.22\% & 80.53\% & 83.50\% & 85.23\%          \\
\textbf{Top-5 CALTeC}        & 89.47\% & 92.00\% & 93.63\% & 94.45\%           \\
\hline
\textbf{Top-1 HaLRTC}        & 77.00\% & 81.75\% & 84.31\% & 85.28\%           \\
\textbf{Top-5 HaLRTC}        & 89.84\% & 92.71\% & 93.94\% & 94.47\%           \\
\hline
\textbf{Top-1 SiLRTC}        & 25.97\% & 32.66\% & 38.62\% & 66.33\%           \\
\textbf{Top-5 SiLRTC}        & 34.32\% & 41.31\% & 47.93\% & 78.07\%           \\
\hline
\textbf{Top-1 Navier-Stokes} & 77.99\% & 81.37\% & 83.88\% & 85.24\%           \\
\textbf{Top-5 Navier-Stokes} & 90.20\% & 92.48\% & 93.75\% & 94.44\% \\
\hline
\end{tabular}
\end{table}

\begin{table}[t]
\centering
\caption{ResNet\nobreakdash-34 \texttt{add\_3} tensors: classification accuracy averaged over 140 Monte Carlo runs per burst loss probability $P_B$}
\label{table:resnet34}
\begin{tabular}{|l|llll|}
\hline
$\boldsymbol{P_B}$      & \textbf{0.3}  & \textbf{0.2}  & \textbf{0.1}   & \textbf{0.01}     \\
\hline
\hline
\textbf{Top-1 NL}   & 85.83\% & 85.83\% & 85.83\% & 85.83\% \\
\textbf{Top-5 NL}   & 94.78\% & 94.78\% & 94.78\% & 94.78\% \\
\hline
\textbf{Top-1 NC}   & 26.27\% & 50.94\% & 74.93\% & 85.58\% \\
\textbf{Top-5 NC}   & 41.77\% & 68.09\% & 87.81\% & 94.62\% \\
\hline
\textbf{Top-1 ALTeC} & 63.24\% & 79.12\% & 84.88\% & 85.96\% \\
\textbf{Top-5 ALTeC} & 77.98\% & 90.20\% & 94.04\% & 94.82\% \\
\hline
\textbf{Top-1 CALTeC} & \textbf{77.31\%} & \textbf{82.86\%} & \textbf{85.35\%} & \textbf{85.99\%} \\
\textbf{Top-5 CALTeC} & \textbf{90.03\%} & \textbf{93.31\%} & \textbf{94.59\%} & \textbf{94.87\%} \\
\hline
\textbf{Top-1 HaLRTC} & 72.84\% & 81.53\% & 85.09\% & 85.95\% \\
\textbf{Top-5 HaLRTC} & 86.01\% & 92.29\% & 94.49\% & 94.85\% \\
\hline
\textbf{Top-1 SiLRTC}  & 57.15\% & 66.21\% & 72.64\% & 83.41\% \\
\textbf{Top-5 SiLRTC}  & 68.41\% & 76.57\% & 82.34\% & 92.79\% \\
\hline
\textbf{Top-1 Navier-Stokes} & 71.45\% & 81.08\% & 84.97\% & 85.98\% \\
\textbf{Top-5 Navier-Stokes} & 85.62\% & 92.04\% & 94.34\% & 94.85\% \\
\hline
\end{tabular}
\end{table}

\begin{table}[t]
\centering
\caption{DenseNet\nobreakdash-121 \texttt{pool2\_conv} tensors: classification accuracy averaged over 140 Monte Carlo runs per burst loss probability $P_B$}
\label{table:densenet121pool2conv}
\begin{tabular}{|l|llll|}
\hline
$\boldsymbol{P_B}$   & \textbf{0.3} & \textbf{0.2} & \textbf{0.1}   & \textbf{0.01}     \\
\hline
\hline
\textbf{Top-1 NL}        & 89.00\% & 89.00\% & 89.00\% & 89.00\% \\
\textbf{Top-5 NL}        & 96.03\% & 96.03\% & 96.03\% & 96.03\% \\
\hline
\textbf{Top-1 NC}            & 42.15\% & 66.84\% & 82.50\% & 88.52\% \\
\textbf{Top-5 NC}            & 56.21\% & 80.60\% & 92.68\% & 95.98\% \\
\hline
\textbf{Top-1 ALTeC}         & 78.93\% & 85.19\% & 87.98\% & \textbf{88.93\%} \\
\textbf{Top-5 ALTeC}         & 91.15\% & 95.08\% & 96.02\% & 96.06\% \\
\hline
\textbf{Top-1 CALTeC}        & 81.94\% & 85.61\% & 87.80\% & 88.89\% \\
\textbf{Top-5 CALTeC}        & 92.79\% & 95.11\% & 96.01\% & 96.05\% \\
\hline
\textbf{Top-1 HaLRTC}        & \textbf{83.38\%} & \textbf{86.72\%} & \textbf{88.32\%} & 88.86\% \\
\textbf{Top-5 HaLRTC}        & \textbf{93.78\%} & \textbf{95.62\%} & \textbf{96.09\%} & \textbf{96.06\%} \\
\hline
\textbf{Top-1 SiLRTC}        & 13.90\% & 19.68\% & 30.16\% & 72.71\% \\
\textbf{Top-5 SiLRTC}        & 18.68\% & 26.04\% & 38.66\% & 82.88\% \\
\hline
\textbf{Top-1 Navier-Stokes} & 82.09\% & 85.87\% & 87.85\% & 88.83\% \\
\textbf{Top-5 Navier-Stokes} & 92.73\% & 95.15\% & 95.98\% & 96.05\% \\
\hline
\end{tabular}
\end{table}

\begin{table}[t]
\centering
\caption{EfficientNet-B0 \texttt{block2b\_add} tensors: classification accuracy averaged over 140 Monte Carlo runs per burst loss probability $P_B$}
\label{table:efficientnetb0}
\begin{tabular}{|l|llll|}
\hline
$\boldsymbol{P_B}$    & \textbf{0.3} & \textbf{0.2}    & \textbf{0.1}   & \textbf{0.01}              \\
\hline
\hline
\textbf{Top-1 NL}         & 86.28\% & 86.28\% & 86.28\% & 86.28\%           \\
\textbf{Top-5 NL}         & 96.15\% & 96.15\% & 96.15\% & 96.15\%           \\
\hline
\textbf{Top-1 NC}         & 0.30\% & 2.80\% & 17.38\% & 75.51\%           \\
\textbf{Top-5 NC}         & 0.69\% & 4.96\% & 25.53\% & 87.98\%           \\
\hline
\textbf{Top-1 ALTeC}  & \textbf{71.64\%} & \textbf{79.64\%} & \textbf{83.89\%} & \textbf{86.05\%}           \\
\textbf{Top-5 ALTeC}  & \textbf{85.06\%} & \textbf{91.31\%} & 94.80\% & 96.08\%           \\
\hline
\textbf{Top-1 CALTeC}   & 65.11\% & 76.06\% & 82.62\% & 85.93\%           \\
\textbf{Top-5 CALTeC}   & 83.55\% & 90.89\% & 94.65\% & \textbf{96.10\%}           \\
\hline
\textbf{Top-1 HaLRTC}  & 65.95\% & 76.69\% & 82.96\% & 85.98\%           \\
\textbf{Top-5 HaLRTC}  & 83.40\% & 91.11\% & \textbf{94.82\%} & 96.09\%           \\
\hline
\textbf{Top-1 SiLRTC}  & 65.71\% & 76.51\% & 82.83\% & 85.97\%           \\
\textbf{Top-5 SiLRTC}  & 83.07\% & 90.88\% & 94.71\% & 96.08\%           \\
\hline
\textbf{Top-1 Navier-Stokes} & 61.14\% & 74.44\% & 82.32\% & 85.91\%           \\
\textbf{Top-5 Navier-Stokes} & 80.34\% & 90.19\% & 94.57\% & 96.12\%           \\
\hline
\end{tabular}
\end{table}

For all DNNs and split layers considered, ALTeC, CALTeC, HaLRTC and Navier-Stokes inpainting bring clear improvements in cloud Top-1 and Top-5 accuracy performance at all loss probabilities $P_B$ over the `No Completion' case. Overall, SilRTC has the worst performance. %While it shows improvement at $P_B=\{0.2,0.3\}$ for the ResNets, for DenseNet\nobreakdash-121 it is significantly worse than the `No Completion' case at all burst loss probabilities. With more iterations, it may be possible to obtain improved SiLRTC results, at the cost of higher inference latency.
While it shows improvements at $P_B = \{0.2,0.3\}$ for ResNet\nobreakdash-18 \texttt{add\_1} and ResNet\nobreakdash-34 \texttt{add\_3} and at all burst loss probabilities for EfficientNet\nobreakdash-B0 \texttt{block2b\_add}, SiLRTC leads to significantly worse performance than the `No Completion/NC' case in all other scenarios. With additional iterations, it may be possible to obtain improved SiLRTC results, at the cost of poorer inference latency.

We present an example of a test image using ResNet\nobreakdash-34 split at the output of \texttt{add\_3}. Fig.~\ref{fig:exp:tiled} shows the test image and tiled versions of the front-end sub-model's original output tensor, the same tensor corrupted after transmission over a GE channel with $(P_B,L_B) = (0.3,7)$ and after repairs by various methods in DFTS2. %50 iterations of HaLRTC and SiLRTC were done on the corrupted tensor. 
When no packet loss concealment was done on the corrupted tensor, the back-end sub-model gave an incorrect Top-1 predicted class with $29\%$ confidence. Meanwhile, all five packet concealment methods produced repaired tensors which led to the correct class (`meerkat') %being predicted by the back-end sub-model 
with high confidence ($>93\%$ with all five methods).
Repaired tensors in Fig.~\ref{fig:exp:tiled} are visually close to the original tensor.

\section{Conclusions} \label{sec:conclusions}
We presented a simulation framework called DFTS2 to study deep feature transmission in collaborative intelligence. We discussed the main components of this simulator, namely, the packetization, communication channel models, and the packet loss concealment schemes. We also presented a comprehensive study of packet loss concealment methods for collaborative image classification. Among these, HaLRTC and CALTeC produced the best Top-1 classification accuracy performance overall in Monte Carlo experiments on three deep neural networks. The proposed simulation framework is flexible and extensible. It can be readily adapted to investigate deep feature compression in collaborative intelligence, as well as the trade-off between compression and error resilience. It can also be easily extended to study other collaborative intelligence and video coding for machines (VCM) scenarios, such as collaborative object detection, segmentation, tracking, and so on.

\clearpage

\bibliographystyle{IEEEbib}
\bibliography{refs}
\end{document}